# Properties of original impactors estimated from three-dimensional analysis of whole Stardust tracks

Michael Greenberg, Denton S. Ebel

Department of Earth and Planetary Sciences, American Museum of Natural History
200 Central Park West, New York, NY 10024

**Abstract –** The Stardust mission captured comet Wild 2 particles in aerogel at 6.1 km/sec. We performed high resolution three-dimensional imaging and X-ray fluorescence mapping of whole cometary tracks in aerogel. We present the results of a survey of track structures using Laser Scanning Confocal Microscopy, including measurements of track volumes, entry hole size and cross-sectional profiles. We compare various methods for measuring track parameters. We demonstrate a methodology for discerning hypervelocity particle ablation rates using synchrotron-based X-ray fluorescence, combined with mass and volume estimates of original impactors derived from measured track properties. Finally, we present a rough framework for reconstruction of original impactor size, and volume of volatilized material, using our measured parameters. The bulk of this work is in direct support of non-destructive analysis and identification of cometary grains in whole tracks, and its eventual application to the reconstruction of the size, shape, porosity and chemical composition of whole Stardust impactors.

## INTRODUCTION

NASA's Stardust mission returned to Earth in 2006 having captured coma dust from comet Wild 2 in an aerogel medium at a relative velocity of 6.1 km/s. Particles captured in aerogel formed 'tracks,' or cavities in the aerogel, where a single track represents the history of one hypervelocity capture event, wherein original impactors were heated, partially volatilized, disaggregated and dispersed (Brownlee et al. 2006, Tsou et al. 2003). The cometary Stardust sample suite contains thousands of these tracks, of varying sizes and morphologies. Burchell et al. (2008) have categorized the shapes of Stardust tracks into three groups: types A, B and C, with shapes resembling carrots, turnips and bulbs respectively. Independent light gas gun experiments in the laboratory have reproduced these three types of tracks, though not necessarily with comet-like material (Hörz et al. 2006, 2008; Burchell et al. 2008). It is clear that track formation is a highly complex process, involving both thermal and physical alteration of particles of all sizes.

Particles extracted from tracks have provided new insights into early solar system processes, although the bulk composition, porosity, and structure of the majority of original impactors is poorly constrained. High-temperature minerals identical to those found in CAIs and chondrules are abundant, indicating either high-efficiency mixing of the protoplanetary disk, or a means of creating localized heating effects (Ciesla 2007; Nakamura et al. 2008). Materials returned by Stardust should give insights to these processes if, and only if, they are properly characterized. Determining the extent of thermal alteration, bulk chemical composition, and original particle bulk physical properties will help to better characterize the cometary source of the Stardust sample suite.

Curation efforts and previous studies of morphological track parameters have utilized conventional optical microscopy limited to two-dimensions (Nakamura-Messenger et al. 2007). Burchell et al. (2006) analyzed two-dimensional images of numerous tracks, deriving a simple relationship between track length, entrance hole diameter, and maximum track width. Our previous work with whole stardust tracks focused on imaging the structures in three-dimensions using simple analysis techniques (Ebel and Rivers 2007, Greenberg and Ebel 2009, 2010). Others investigating the structure of tracks in three dimensions include Kearsley et al. (2007), Tsuchiyama et al. (2009) and Iida et al. (2010).

**Table 1:** Track analysis conditions reported in this study.

| Track | Curatorial Name | Type[a] | LSCM X-Y Sampling ($\mu m$) | Z Sampling ($\mu m$) | Voxel Dwell ($\mu s$) | Dataset Size (GB) | SXRF Energy (KeV) | Resolution ($\mu m$) | Dwell Time (s) | Number of Submaps |
|---|---|---|---|---|---|---|---|---|---|---|
| 82 | C2092,1,082,0,0 | A | 0.075 - 0.225 | 0.34 -0.824 | 12.8 | 2.7 | 21 | 2 | 2 | 1 |
| 128α | C2012,4,128,0,0 | A | 0.075 | 0.824 | 6.4 | 0.67 | 21 | 2 | 2 | 2 |
| 128β | C2012,4,128,0,0 | A | 0.075 | 0.824 | 6.4 | 0.38 | 21 | [c] | [c] | [c] |
| 128γ | C2012,4,128,0,0 | A | 0.45 | 0.824 | 0.6 | 0.05 | 21 | [c] | [c] | [c] |
| 128δ | C2012,4,128,0,0 | A | 0.037 | 0.312 | 0.8 | 0.05 | 21 | [c] | [c] | [c] |
| 129 | C2012,5,129,0,0 | A | 0.075 | 0.5 | 6.4 | 1.24 | 21 | [c] | [c] | [c] |
| 140 | C2061,2,140,0,0 | B | [b] | [b] | [b] | [b] | 21 | 10 | 0.5 | 5 |
| 151 | C2112,1,151,0,0 | A | 0.318 | 0.838 | 1.6 | 30 | 21 | 5, 10 | 0.5 | 9 |
| 152 | C2035,2,152,0,0 | A | 0.083 | 0.359 | 1.6 | 53.4 | 21 | 5 | 0.5 | 11 |
| 163 | C2117,1,163,0,0 | B | 0.083 | 0.359 | 3.2 | 33.2 | 21 | 1 | 0.1 | 2 |
| 164 | C2063,1,164,0,0 | B | 0.359 | 0.7 | 12.8 | 2.3 | 21 | 1 | 0.1 | 2 |

[a] Classified from Hörz et al. (2006)

[b] Sample not imaged with LSCM

[c] Sample not analyzed with SXRF

It has been our focus to chemically and morphologically characterize whole tracks in three dimensions, utilizing solely nondestructive methods. Recently, we reported on the results from a variety of methods: three-dimensional Laser Scanning Confocal Microscopy (LSCM), synchrotron X-ray fluorescence (SXRF), synchrotron X-ray diffraction (SXRD), and synchrotron radiation X-ray microtomography (SR-XRμCT). This study reports the results of three-dimensional imaging of track structures using LSCM, in combination with high-resolution SXRF chemical mapping. The combination of these techniques is used to obtain quantitative track morphological data, in conjunction with inferences of mass ablation for impacting particles along tracks (Anderson and Ahrens 1994, Dominguez et al. 2004, 2009; Coulson 2009; Price et al.

2010). This work is intended as a preliminary roadmap for how nondestructive three-dimensional data can be put to use by Stardust researchers to better understand the properties of the entire Stardust sample suite. The estimates of properties of whole Stardust impactors, presented in the final section of this paper, are intended as first-order estimates of how fully nondestructive, three-dimensional data on whole Stardust tracks can be combined with current hypervelocity impact modeling to more thoroughly understand the original impactors from Stardust. While it remains unclear exactly how the myriad of shapes and fine scale structures of Stardust tracks evolve and form, our data provides clues and high-resolution, ground-truth observations to those stustudying hypervelocity impacts in aerogel from a theoretical and experimental standpoint.

## METHODS

Eight keystones (Westphal et al. 2002, 2004), containing a total of eleven cometary tracks, were analyzed in this study. Two of these tracks have been partially reported on in previous studies. We have chosen not to nickname tracks, but will refer to segments of their curatorial name as follows: track 82 (C2092,1,82,0,0); keystone 128 (C2012,4,128,0,0), which has four tracks that are referred to as 128α, 128β, 128γ, and 128δ respectively; track 129 (C2012,5,129,0,0) ; track 140 (C2061,2,140,0,0), previously reported in Tsuchiyama et al. 2009, nicknamed Pineapple) ; track 151 (C2112,1,151,0,0) ; track 152 (C2035,2,152,0,0) ; track 163 (C2117,1,163,0,0) ; and track 164 (C2063,1,164,0,0). Table 1 summarizes the tracks and the analysis methods applied in this study.

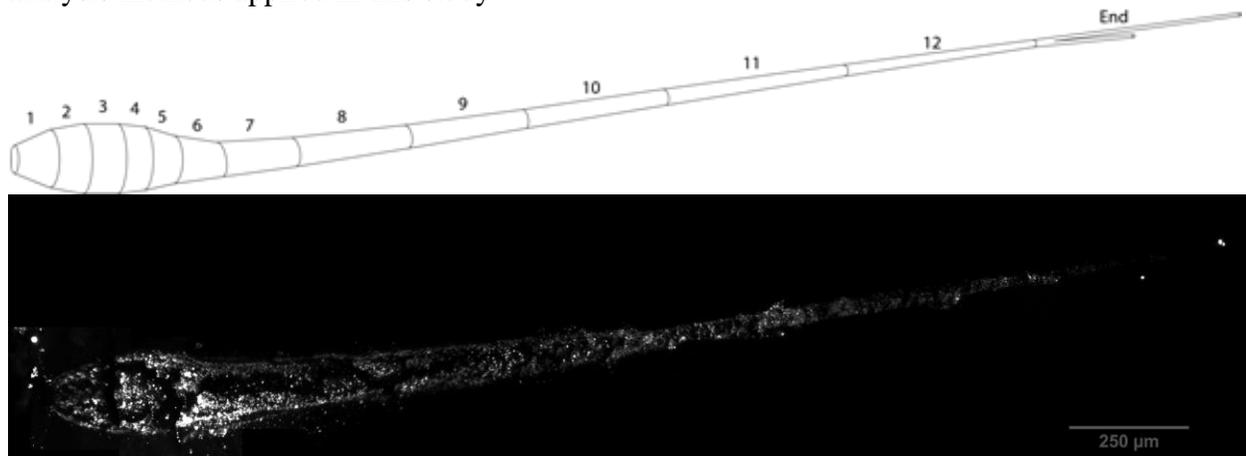

**Figure 1**: (Top) Cartoon of frustum divisions used to calculate morphology parameters on Track 152. (Bottom) LSCM reflectance intensity map, 8-bit grayscale image in line with frustum cartoon.

The three-dimensional imaging of these tracks was performed by Laser Scanning Confocal Microscopy (LSCM) at the Microscopy and Imaging Facility at the American Museum of Natural History in New York City. Previous iterations of our LSCM imaging technique have been described elsewhere (Greenberg and Ebel, 2010a, b, c), and the images presented here represent over two years of development of those techniques. The LSCM used in this study is a Zeiss LSM 510 (c. 1997), an inverted type LSCM on an Axiovert 100 mount. All images presented use 488 nm laser illumination in reflectance mode. None of the fluorescence capabilities of the LSM 510 were used in this study. Using the LSM 510, two-dimensional image "slices" are acquired by varying the distance between a series of lenses and the sample. An adjustable pinhole blocks varying amounts of light from adjacent planes. Smaller pinhole diameters can be used to block increasing amounts of light, creating thinner optical slices, but with a loss in reflectance intensity. As a guideline, a 1 Airy unit pinhole size was adopted for all images. This procedure is used to create image "stacks" as a three-dimensional representation of the structure being imaged. The LSM 510 captures images on a 2048 x 2048 pixel CCD that can be binned appropriately to create lower resolution imagery. Our suite of images was acquired at proper Nyquist sampling rates when reasonable within time constraints (typically under a week; the LSM 510 is a shared instrument). Image sets can have hundreds of image slices and imagery of whole tracks is typically comprised of several image stacks (at proper Nyquist sampling rates, a 20x objective has a field of view of ~170 x 170 µm). Due to diffraction of light around the pinhole aperture, sampling in the Z direction (optic axis of the microscope) is much more coarse than in the X and Y directions. This phenomenon is inherent to all images produced by LSCM. The axial distortion in the Z direction can be partially corrected in post-processing; these techniques have been described elsewhere (Greenberg and Ebel 2010c). Table 1 summarizes the LSCM image datasets used in this study.

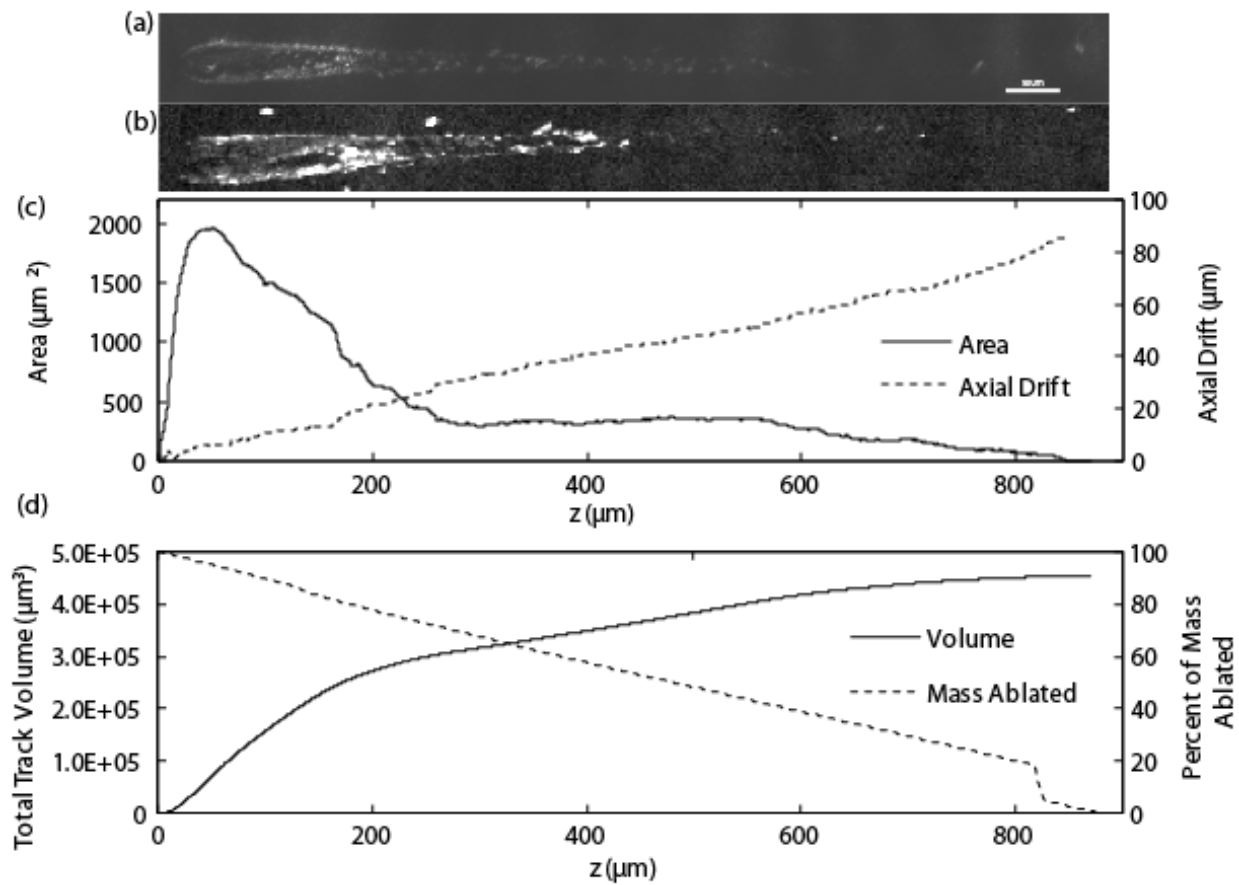

**Figure 2**: Morphology and composition results for Track 82, all images are vertically aligned (**a**) LSCM map of Track 82, single plane at 8-bit depth, a contrast stretch has been applied to increase visual contrast. (**b**) SXRF Fe Kα fluorescence map, track 82, large bright spots above the track are surface aerogel contaminants. (**c**) Cross sectional area and axial drift versus penetration depth (z). (**d**) Track volume and percentage of impactor mass vs. penetration depth (z), terminal particle is included in mass ablation data.

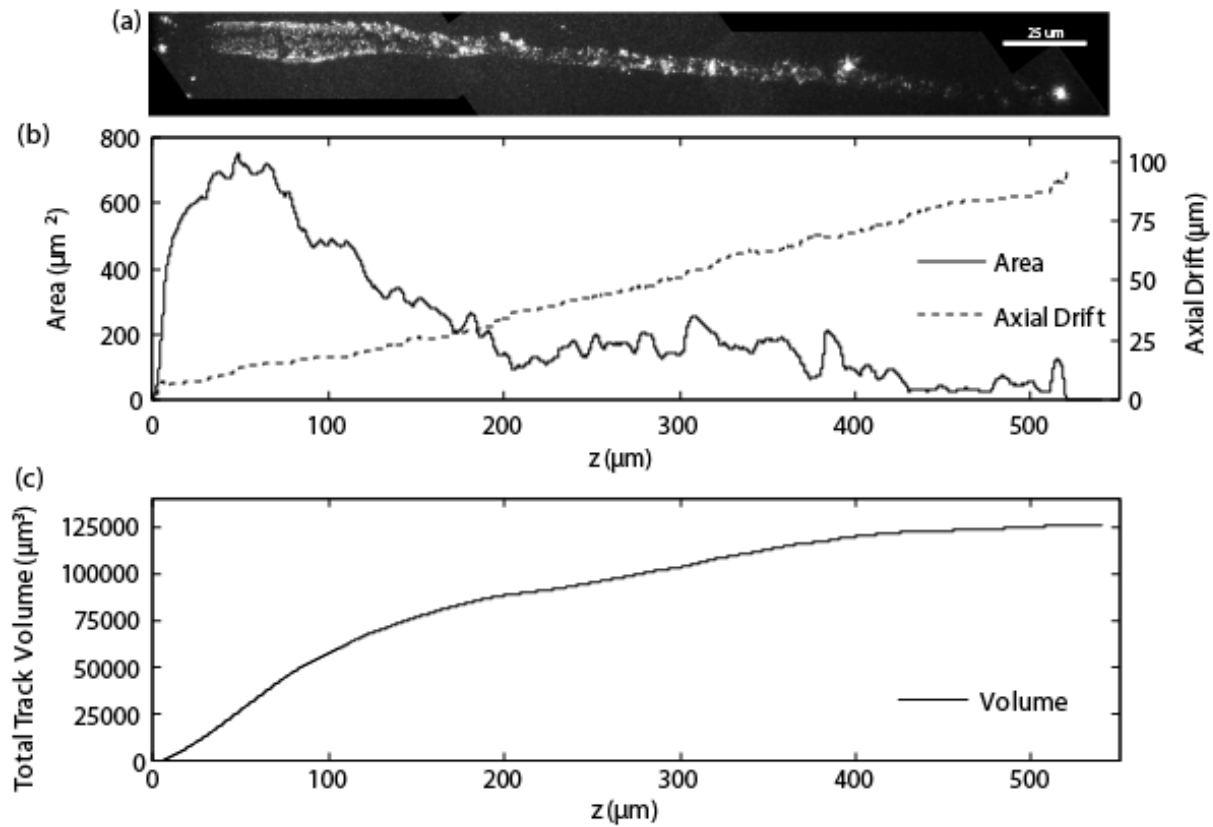

**Figure 3**: Morphology and composition results for Track 128α, all images are vertically aligned (**a**) LSCM map of Track 128α, single plane at 8-bit depth, a contrast stretch has been applied to increase visual contrast. (**b**) Cross sectional area and axial drift versus penetration depth (z). (**c**) Track volume versus penetration depth (z). SXRF data on track 128α is inconclusive.

    Multiple image stacks are stitched together in post processing to create monolithic three-dimensional datasets for image analysis. Track images were manually binarized using NIH ImageJ v1.45. Manual binarization consisted of assigning two specific grayscale values to two areas of each image, one grayscale value for areas inside the aerogel cavity, and the other grayscale value for areas outside the aerogel cavity; this results in a binarized, or segmented image. Automated segmentation of confocal imagery is not yet feasible, due to the complexities and variance in track structures. Morphology calculations are the results of processing on binarized image stacks using Python v.3.1 scripts.

    For comparison with full three-dimensional binarized images, two other binarization approaches were used. 1) A single, maximum intensity projection of a whole LSCM dataset was binarized by hand to estimate errors induced by downsampling three-dimensional data to two-dimensions. This single outline was used to calculate track volume by integrating sequential disks as volumes of rotation orthogonal to the penetration direction. 2) A piecewise frustum (vertical segments of cones) method (Fig. 1a) was used to approximate track size and orientation in three-dimensions. In the frustum method tracks were divided by visual inspection into multiple frustums, with divisions occurring typically once every 100 μm (sizes vary based on track size and morphology). The bases of the frustums were measured by binarization of resliced LSCM data orthogonal to the direction of impactor motion, and parameters were quantified using

a five-slice average. Figures 1 through 9 display morphology data including maximum intensity projections of LSCM data, and single outlines of tracks.

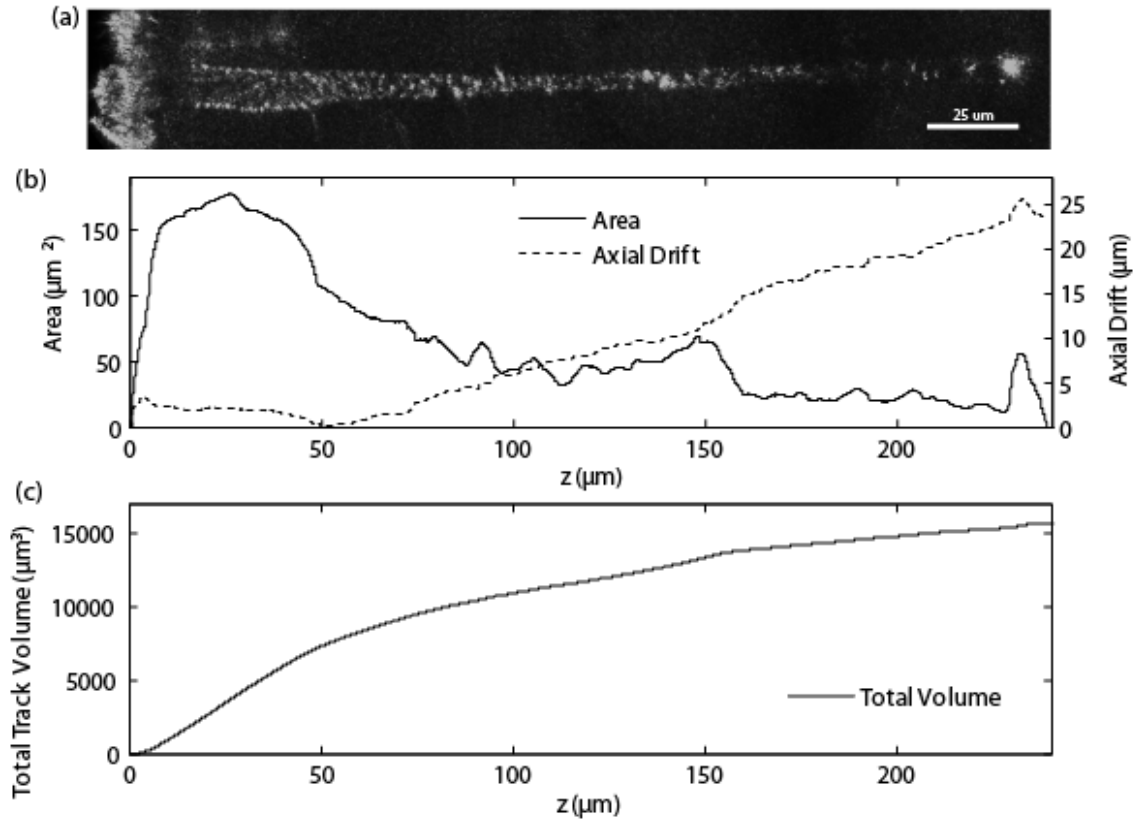

**Figure 4**: Morphology and composition results for Track 128β, all images are vertically aligned (**a**) LSCM map of Track 128β, single plane at 8-bit depth, a contrast stretch has been applied to increase visual contrast. (**b**) Cross sectional area and axial drift vs. penetration depth (z). (**c**) Track volume versus penetration depth (z). SXRF data on track 128β is inconclusive.

Chemical compositions of tracks 82, 128α, 140, 151, 152, 163 and 164 were determined using synchrotron X-ray fluorescence techniques, over the course of four separate synchrotron runs (between 2007-2010). SXRF data were collected at the Advanced Photon Source (APS) at Argonne National Labs (ANL), beamline 13-ID-C (GeoSoilEnviroCars). At the APS, a pair of Kirkpatrick-Baez microfocusing mirrors are used in tandem with a Si (111) monochromator to focus the incident 300 x 300 μm X-ray beam to a 3 x 4 μm spot, with the incident energy held constant at 21 KeV. Spectra are collected with a 4-channel Vortex silicon drift detector and are calibrated using NIST 1832 and 1833 SRM standards (Sutton 2006). Maps are collected by rastering the sample in two dimensions through the beam with dwell times ranging from 0.1 – 2s per spot, with step sizes ranging from 1 – 10 μm, depending on the size of the sample. Improvements in the experimental setup at APS over the course of this study have significantly reduced data collection times by an order of magnitude. In previous studies we have reported SXRF results by binning disparate areas of track by visual inspection. We have also reported the results of long (60 – 180 s) dwells on bright spots, in a method similar to other SXRF studies of Stardust material (Flynn et al. 2006, Ishii et al. 2008, Lanzirotti et al. 2008, Ebel et al. 2009).

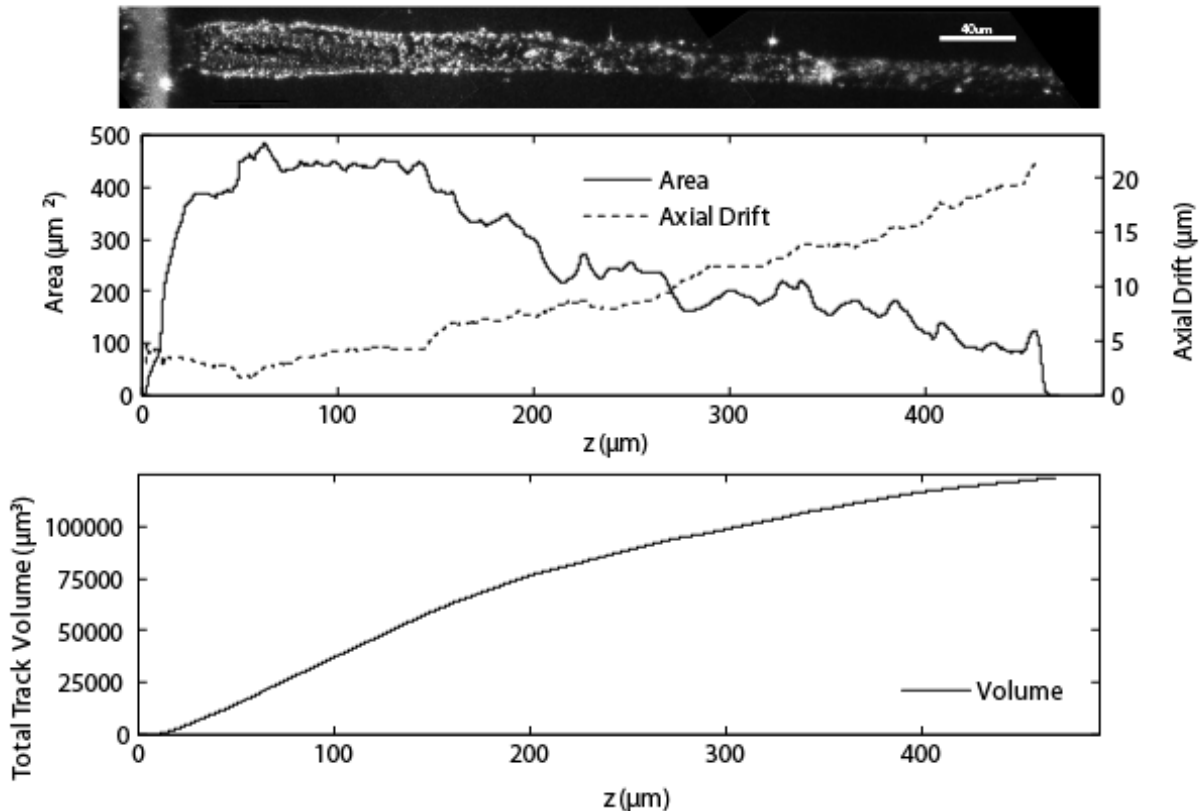

**Figure 5**: Morphology and composition results for Track 129, all images are vertically aligned (**a**) LSCM map of Track 129, single plane at 8-bit depth, a contrast stretch has been applied to increase visual contrast. (**b**) Cross sectional area and axial drift vs. penetration depth (z). (**c**) Track volume versus penetration depth (z). No SXRF data exists for track 129.

Our goals for this manuscript are slightly different. We wish to estimate particle ablation rates from SXRF maps. To estimate total particle mass along tracks, SXRF results were scaled to CI chondritic abundances (Lodders 2003) from measured intensities of the Fe K$\alpha$ fluorescence (e.g., Flynn et al. 2006). While the assumption of CI bulk composition cannot be correct (Flynn et al. 2006), it renders our work consistent with earlier work using the same assumption. Un-normalized SXRF results yield relative track abundances only for elements with $Z > 16$. In this work we estimate particle ablation rates by integrating lines of SXRF map data, orthogonal to track direction. A spot background subtraction correction is calculated by binning large portions of blank aerogel spectra and scaling to the length of each row. Background correction, spot size corrections and absorption corrections are all applied before summing entire maps to determine total track mass. Ablation profiles are acquired by subtracting subsequent rows from total integrated mass.

Synchrotron X-ray diffraction was also attempted on tracks 140, 151, and 152 at the APS and at Brookhaven National Labs (beamline X26A), but the data yielded no useful results. This was due to very small grain sizes and poor signal/noise ratios in the results, coupled with a reluctance to subject samples to high X-ray flux with long dwell times.

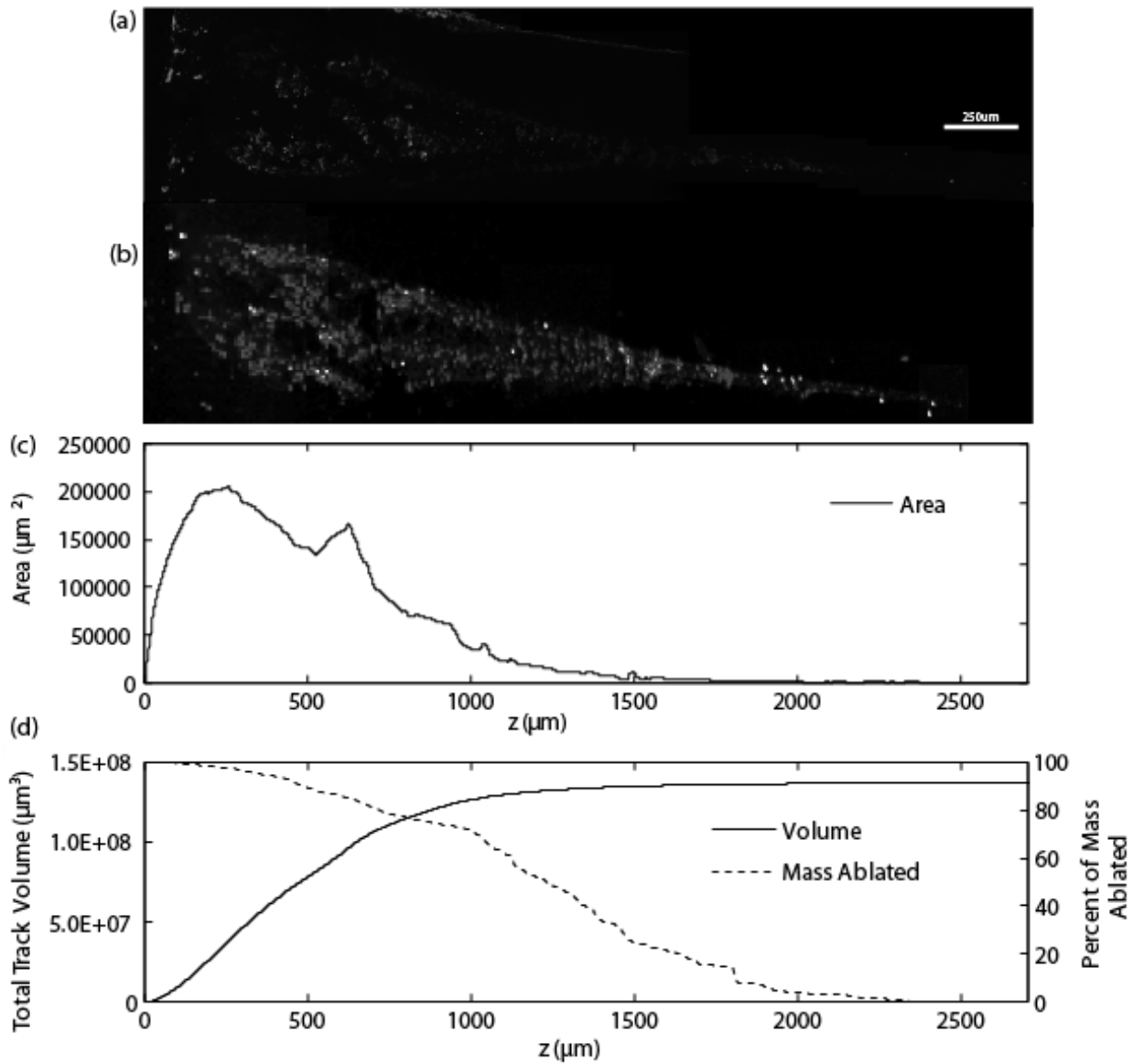

**Figure 6:** Morphology and composition results for Track 151, all images are vertically aligned (**a**) LSCM map of Track 151, eight planes at 12-bit depth, a contrast stretch has been applied to increase visual contrast. (**b**) SXRF Fe Kα fluorescence map, track 151 (**c**) Cross sectional area vs. penetration depth (z). (**d**) Track volume and percentage of impactor mass versus penetration depth (z), terminal particle is not included in mass ablation data. Morphology parameters calculated via single slice method.

## RESULTS

Automated segmentation techniques cannot be applied to our image stacks due to the complexity of track structure. Unlike in computed tomography (CT) images (Nakamura et al.

2008, Tsuchiyama et al. 2009), in LSCM imagery there is no one grayscale value which can be used to define the boundary between track walls and cavity space, and threshold values may vary both laterally and by depth due to laser attenuation. For optimal results, all data are manually segmented. Binarized images are rotated to a vertical alignment and are resliced orthogonal to impact direction. Python scripts are then used to calculate a variety of track parameters.

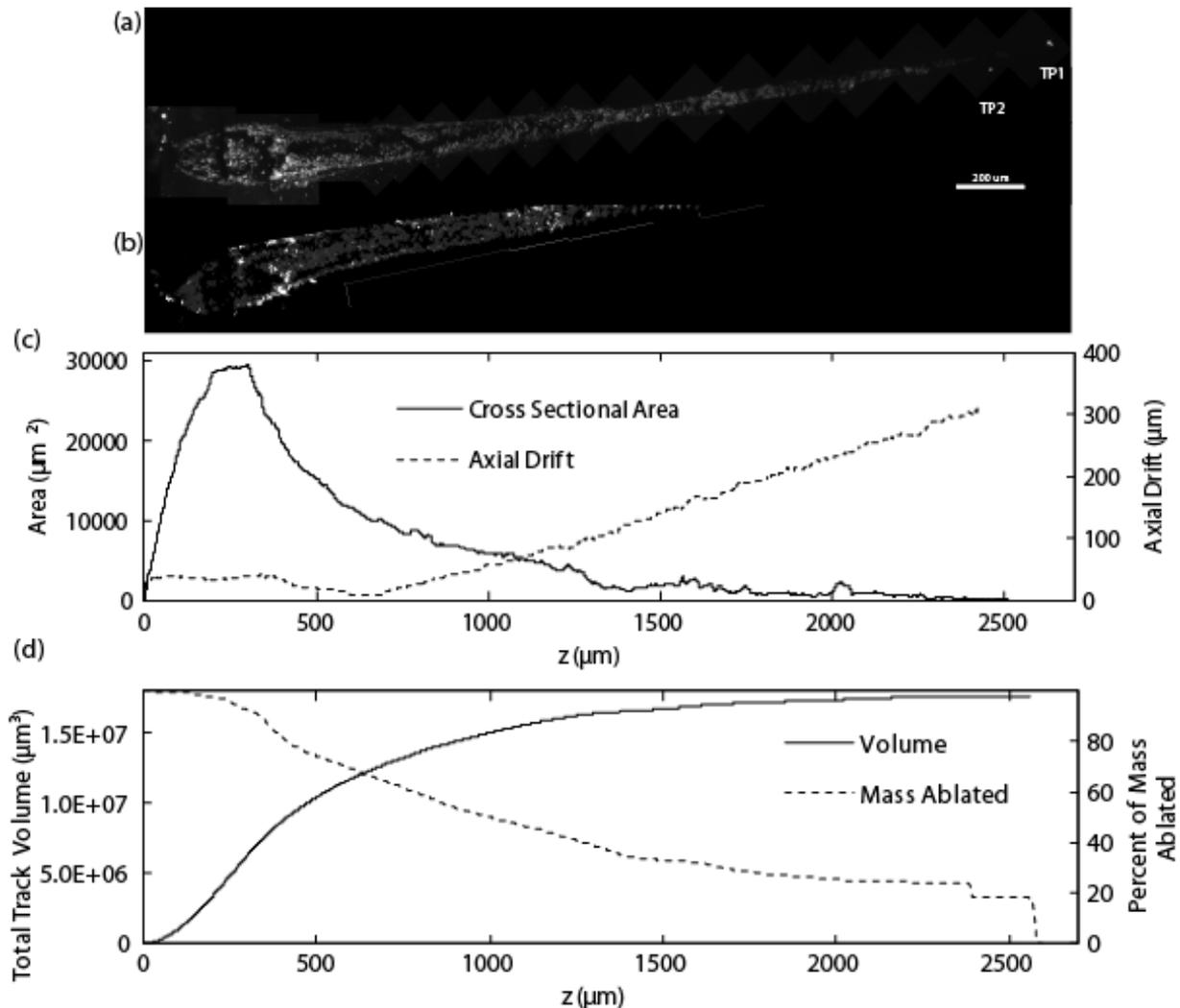

**Figure 7:** Morphology and composition results for Track 152, all images are vertically aligned (**a**) LSCM map of Track 152, 18 planes at 12-bit depth, a contrast stretch has been applied to increase visual contrast. (**b**) SXRF Fe K$\alpha$ fluorescence map, track 152. (**c**) Cross sectional area and axial drift vs. penetration depth (z). (**d**) Track volume and percentage of impactor mass versus penetration depth (z), terminal particle is included in mass ablation data.

For each slice orthogonal to the track axis, a cross sectional area is calculated, and a Gaussian ellipse is fitted (Gander et al. 1994), from which major and minor axes, eccentricity and a rotational orientation can be derived. Local minimum and maximum of cross-sectional area at different depths are adopted as the entrance hole diameter and the maximum track width. Additionally, the data can be used to calculate track volumes, and cumulative volume. Furthermore, we define a parameter called axial deviation, which quantifies the deviation of the centroid of any given slice from a normal (true 0º) impact. A track with an impact angle of 0º will have a total axial drift of zero. This parameter can be used to quantify track motion in three-dimensions and any fine scale angular shifts. The total axial deflection is used to calculate the impact angle parameter. This data is presented as a function of D, the penetration of impactor depth, or distance from the entrance hole. The location of the entrance hole may differ from the actual top surface of the aerogel, but we have adopted this method to stay in line with the measurements of others.

**Table 2:** Track parameters reported in this study.

| Track | Curatorial Name | Type | Length (µm) | Terminal Particles | Volume (µm$^3$) | Entrance Hole Area (µm$^2$) | Entrance Hole Eccentricity | Track Maximum Area (µm$^2$) | Track Maximum Eccentricity | Total Axial Drift (µm) | Impact angle (degrees) | Total Fe Mass in Track (fg) |
|---|---|---|---|---|---|---|---|---|---|---|---|---|
| 82 | C2092,1,082,0,0 | A | 898 | 1 | 4.53E+05 | 1036.3 | 0.46 | 2219.2 | 0.74 | 85.5 | 9.5 | 3.76E+04 |
| 128α | C2012,4,128,0,0 | A | 520 | 2 | 1.26E+05 | 325.9 | 0.71 | 652.4 | 0.62 | 97.4 | 18.8 | c |
| 128β | C2012,4,128,0,0 | A | 250 | 1 | 1.57E+04 | 178 | 0.68 | 199.9 | 0.79 | 23.7 | 9.5 | c |
| 128γ | C2012,4,128,0,0 | A | 43 | 1 | 1.28E+02 | 1.1 | e | 2.5 | e | 1.5 | 3.5 | c |
| 128δ | C2012,4,128,0,0 | A | 30 | 1 | 3.61E+02 | 8.5 | e | 8.5 | e | 1.3 | 4.3 | c |
| 129 | C2012,5,129,0,0 | A | 700[a] | 2[d] | 1.23E+05 | 274.1 | 0.36 | 649.3 | 0.61 | 25.6 | 3.7 | c |
| 140 | C2061,2,140,0,0 | B | 4285[b] | 2 | 4.02E+09[b] | 3185.6 | e | 9068524.5 | e | e | e | 1.80E+07 |
| 151 | C2112,1,151,0,0 | A | 3121 | 4 | 1.37E+08 | 1553.7 | e | 205204.7 | e | e | e | 1.37E+07 |
| 152 | C2035,2,152,0,0 | A | 2525 | 3 | 1.76E+07 | 3074.3 | 0.39 | 25127.5 | 0.44 | 425.9 | 16.9 | 7.16E+06 |
| 163 | C2117,1,163,0,0 | B | 1284 | 2 | 2.20E+08 | 3554.3 | 0.46 | 35089.6 | 0.67 | 84.8 | 6.6 | 1.48E+07 |
| 164 | C2063,1,164,0,0 | B | 541 | 3 | 4.33E+06 | 1679.5 | 0.51 | 23632.4 | 0.45 | 34.1 | 6.3 | 1.23E+05 |

[a] Track is incomplete

[b] Tsuchiyama et al. (2009)

[c] SXRF not measured

[d] One terminal particle remains

[e] Unable to calculate

*Imaging Results*

Table 2 lists parameters measured in SXRF and LSCM of all the tracks in this study. Figures 2 - 9 illustrate major quantifiable data on all tracks presented in this work. To maintain consistency, all tracks are presented in the same manner (if each component is available). The topmost image is a maximum intensity projection of the LSCM map, below is a corresponding SXRF map illustrating Fe Kα fluorescence intensity. The two maps are aligned with the subsequent plots, the first of which shows cross sectional area and axial drift as a function of depth. These are plotted on separate axes. The final component plots percentage of impactor ablated and total track volume, each as a function of track depth. These composites are presented in this order so as to create a data structure that can be used to interpret the data.

*Track 82 (C2092,1,82,0,0)*

This keystone contains a single 898µm long track and is mounted on a standard fork and 25mm glass rod. Its three-dimensional geometry (See Figure 2) indicates a spiraling impactor, with a large area of particulate deposition about halfway down the track, coinciding with a noticeable kink in track structure. This track has been previously reported on in extensive detail (Ebel et al. 2008).

*Track(s) 128 (C2012,4,128,0,0)*

This keystone contains four separate tracks of lengths 520, 250, 43, and 30 µm. These have been designated alpha, beta, gamma and delta, respectively. This keystone was mounted on a standard fork until mounting for SXRF analysis, at which point the keystone became detached from its fork. Keystone #128 is now housed in a kapton box, suitable for SXRF analysis. This keystone was imaged using LSCM before being housed in kapton.

Morphologically, T128α is curious, exhibiting spiral behavior as well as small amounts of particulate deposition along most of the track (Figure 3a). There is a small bifurcation present about a third of the way down the track with a small secondary terminal particle in that region. Additionally, there is a terminal particle of high interest about three-quarters down the track, which is slightly smaller than the terminal particle. The bulbous region of T128α does not strictly conform to the standard carrot shape; the cross-sectional area data indicates a series of plateaus, rather than a typical, smooth tapering of the track. Analysis of T128α using SXRF was inconclusive; very low levels of Fe and other detectable elements, barely above background, led to unconvincing measurements. This was probably due to detector failure. Full three-dimensional imagery of T128α is viewable online through the CATMAID framework (Saalfeld et al. 2009) at: bit.ly/hBjkEZ.

Track 128β is a carrot shaped track, which also exhibits some spiraling behavior despite being much less skewed than T128α (Figure 4a). Additionally, there are no secondary terminal particles to be found in T128β, and its trajectory is oriented away from Track 128α, indicating these 2 impacts probably did not form from some clustering event. SXRF data on T128β was also inconclusive, so at this time the bulk chemical composition of both T128α and T128β is unknown. The entry points of T128α and T128β are about 150µm apart and from visual inspection, have largely differing deposition patterns.

T128γ and T128δ are small and thin carrot shaped tracks within the same keystone, each with single terminal particles. Due to their miniscule size, we are unable to detect fine-scale structure in these two tracks. Tracks of this size are difficult to locate in three dimensions with a 20x objective, thus there may be similar tracks in this and other keystones that may have been missed. Images of these tracks have been presented elsewhere (Greenberg and Ebel, 2009) This opportunity presents a previously untapped source of Stardust particles, although their small size may result in increased thermal alteration effects.

*Track 129 (C2012,5,129,0,0)*

This keystone contains one, incomplete, 700µm track, and has been imaged using LSCM (Figure 5a). Morphologically, T129 is very cylindrical and straight in shape, with a small secondary terminal particle present about halfway down the track, creating a small region of axial drift. The cross-sectional area data shows a very linear decrease, and the axial drift data indicates an increasing rate of drift as a function of penetration depth. During transit from Johnson Space Center the tip of this keystone, including the largest terminal particle, broke off, thus this track is missing an estimated 300µm, so total drift is unknown. This keystone is still on a forklift mount.

*Track 140 (C2061,2,140,0,0)*

This keystone houses one bulbous track which is significantly larger than any track we have imaged in the past (>4000 µm in length). Use of lower magnification lenses is required to image this track with LSCM, yet it is easy to see radial fragmentation extending beyond the keystone, and a large terminal particle region non-orthogonal to the plane of incidence. The thickness of this keystone (>1 mm) is greater than the standard penetration depth for a confocal microscope (~600 µm) (Denk et al. 1995). The bulbous portion of this track is ~4100 µm in length and has 2 major styli, each with one large terminal particle at its end. Smaller styli can be seen extending radially outward from the bulb area, but are much shorter in length and have much smaller terminal particles. Unlike previous tracks, keystone 140 has not yet been quantitatively analyzed, but previous work by Tsuchiyama et al. (2009) allows for comparison of morphology to SXRF measurements. Analyses of background elements (e.g. Br) in aerogel indicate uneven compaction of aerogel, and uneven distribution of particles extending radially around the bulb. Due to the two-dimensional nature of SXRF mapping, it is difficult to correlate Fe hot spots with actual particles in three dimensions. This keystone is housed in a rigid kapton sleeve.

*Track 151 (C2112,1,151,0,0)*

This keystone contains one 3121 µm long type A track, with a small bulbous portion extending approximately 1mm down the track (Figure 6a). The bulb is less defined than those of other tracks and gradually forms a stylus. Track 151 has four major terminal particles. Upon transport from NASA Johnson Space Center the keystone housing Track 151 fell off its forklift and was slightly damaged, most noticeably in the bulbous portion. We have housed this track in a rigid kapton sleeve. All analyses have been performed while the keystone is in this housing. Due to the deformation of this track and its keystone, no three-dimensional analysis was done on track 151. Instead, morphological measurements from this track were performed using only the maximum intensity projection method described above.

*Track 152 (C2035,2,152,0,0)*

This keystone contains one 2525 µm long type A track with a small (approximately 315 µm long) bulbous section in the beginning of the track (Figure 7a). Track 152 has three terminal particles; the largest two reside at the ends of the styli, and a smaller particle is found at a penetration depth of about 1700 µm. Combining the morphological data from confocal imagery with elemental data from SXRF provides unparalleled insight into track formation impact events. The initial impactor pulled inwards on the aerogel surface upon impact, creating a dish shaped dimple with a depth of 41 µm, and a radius of 112 µm. The initial track opening is 80 µm in width but quickly closes to a minimum width of 45 µm. At a depth of 135 µm a large radial crack appears near the top of the keystone, and the next radial crack appears 130 µm from that depth, an aerogel fragmentation pattern similar to those produced by hypervelocity analog shots. Maximum bulb width is achieved between the two major radial cracks, and at a depth of 205 µm. In the bulbous area spatial Zn concentration correlates directly with Cu but not with Fe, the typical marker element for SXRF. At a depth of approximately 345 µm there is a large deposition of material (mostly Fe rich) and a dramatic skewing of the track ~18° off normal. The first 700 µm of the skewed portion of the track is marked by a large amount of smaller particulate material deposited radially and one large, mostly continuous, radial fragment. The cross-sectional profile in this region is also markedly non-circular. The region prior to the terminal particles is marked by a generally linear decrease in cross sectional area, but deposition in this region appears periodic in the SXRF map. The two largest terminal particles are at the end of the track, the smaller one is roughly spherical and ~6 µm in diameter, and it is enriched in Cu but depleted in Ni and Cr, relative to the CI norm. The largest of the terminal particles is bifurcated; the smaller fragment is roughly spherical and is ~5 µm in diameter, while the larger fragment is oblong with a length of ~9.5 µm and a width of ~6.5 µm. The bulk composition of this particle is also enriched in Cu and depleted in Ni and Cr, relative to the CI norm. An animation of the entire LSCM dataset on Track 152 is viewable online at: bit.ly/Track152.

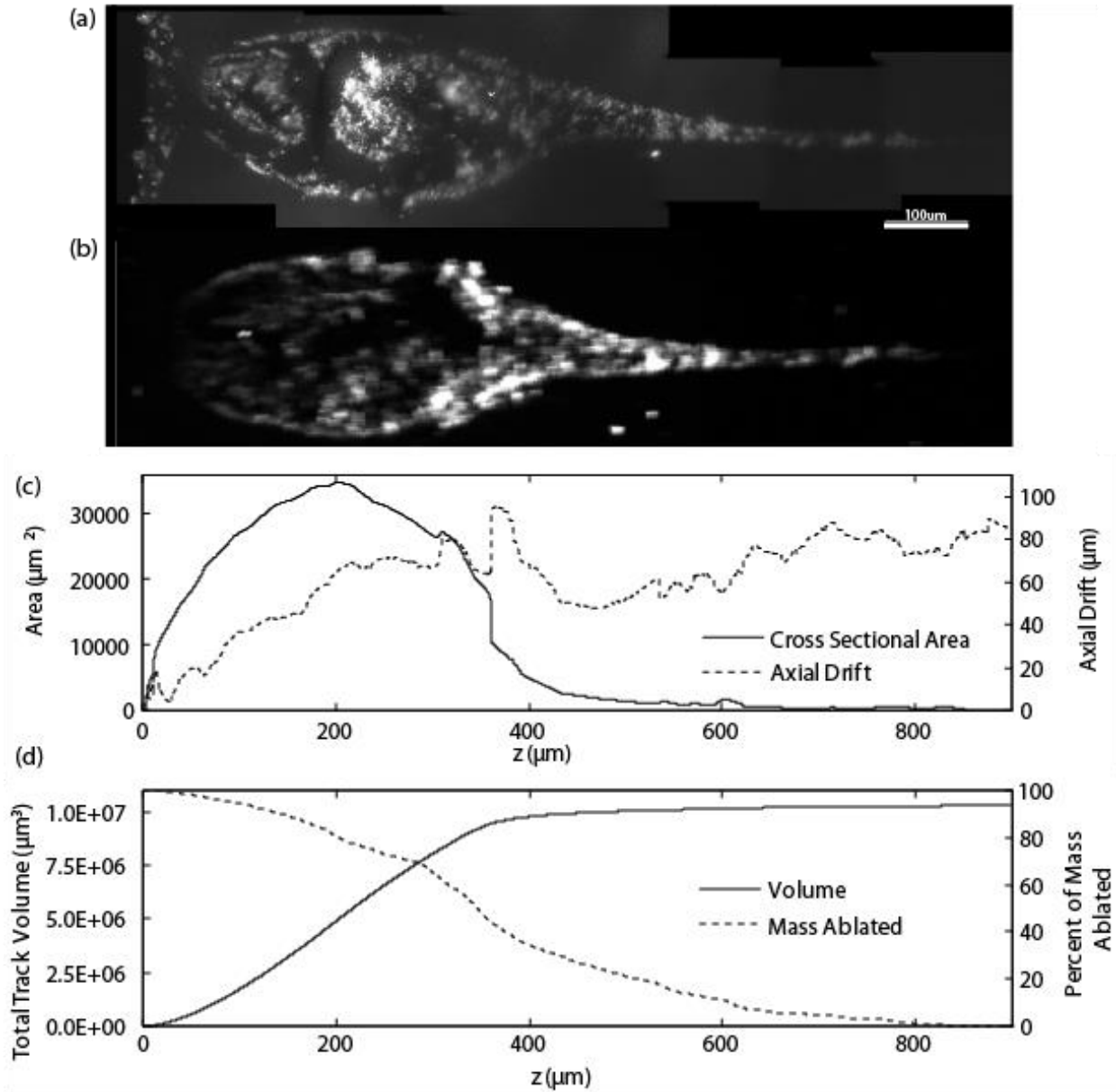

**Figure 8:** Morphology and composition results for Track 163, all images are vertically aligned (**a**) LSCM map of Track 163, 9 planes at 12-bit depth, a contrast stretch has been applied to increase visual contrast. (**b**) SXRF Fe Kα fluorescence map, track 163. (**c**) Cross sectional area and axial drift vs. penetration depth (z). (**d**) Track volume and percentage of impactor mass vs. penetration depth (z), terminal particle is not included in mass ablation data.

*Track 163 (C2117,1,163,0,0)*

This keystone contains one 1284 µm long type B track with a large (~487 µm long, 210 µm wide) bulbous section in the beginning of the track (Figure 8a). Track 163 has two terminal particles: the largest (3 µm diameter) terminal particle residing at the end of the track and a slightly smaller particle residing in a stylus ~613 µm from the entrance hole. Large cracks in aerogel are seen in the bulb area, and the main stylus is 8 degrees askew from the axis of the bulb.

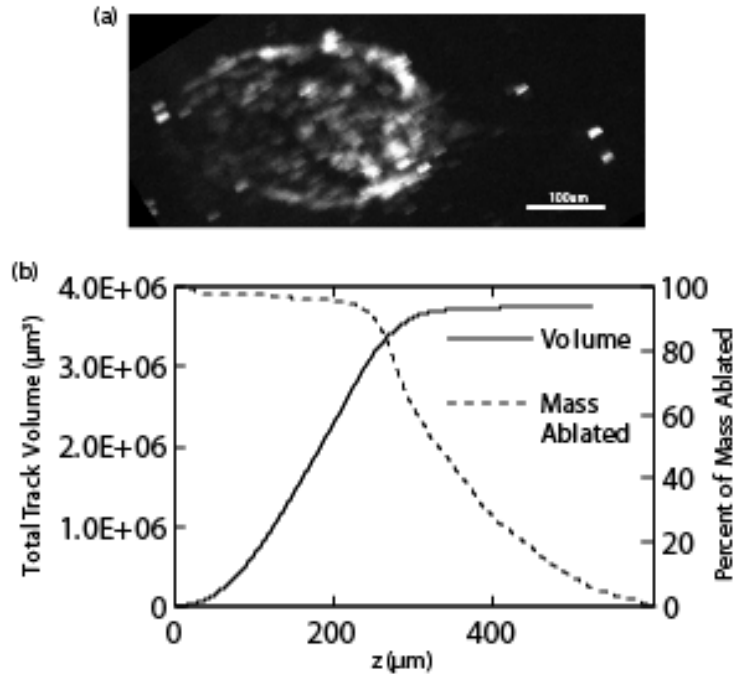

**Figure 9:** Morphology and composition results for Track 164. (**a**) SXRF Fe Kα fluorescence map, track 164. (**b**) Track volume and percentage of impactor mass vs. penetration depth (z), terminal particle is not included in mass ablation data. Track morphology is calculated via single outline method.

*Track 164 (C2063,1,164,0,0)*

This keystone contains one 541 µm long type B track with a small (~346 µm long, 148 µm wide) bulbous section in the beginning of the track (Figure 9a). Track 164 has two terminal particles, both residing at the end of the track, and both ~4 µm in diameter. In transit from JSC, the keystone containing Track 164 fell off its forklift mount, and the keystone was remounted in an envelope of 4 µm thick kapton for analyses and to ensure sample safety.

**Table 3:** Results of three methods for measuring track volume.

| Track | Volume Measurements (µm$^3$) | | | Percent Difference | |
|---|---|---|---|---|---|
| | 3D Outlines | Frustum | 2D Outline | Frustum | 2D Outline |
| 82 | 4.53E+05 | 4.20E+05 | 5.31E+05 | 7.35% | -17.29% |
| 128α | 1.26E+05 | 1.02E+05 | 6.66E+04 | 19.23% | 47.23% |
| 128β | 1.57E+04 | 1.52E+04 | 6.36E+03 | 3.10% | 59.54% |
| 129 | 1.26E+05 | 1.28E+05 | 1.36E+05 | -2.35% | -8.72% |
| 152 | 1.76E+07 | 1.75E+07 | 1.65E+07 | 0.91% | 6.28% |
| 163 | 1.03E+07 | 9.65E+06 | 8.80E+06 | 6.14% | 14.42% |

## DISCUSSION

**Two-Dimensional & Three-Dimensional Volume Measurements**

　　　Full segmentation of three-dimensional LSCM images can be time consuming, even on smaller datasets. As our data cubes grow larger with each iteration of our imaging procedure, it is imperative to develop fast methods for three-dimensional structural analysis. We have made measurements of several tracks with three different methods to demonstrate the potential inaccuracy of two-dimensional imaging for measuring three-dimensional parameters of tracks. Certain tracks were not analyzed due to a lack of three-dimensional images. We calculated volume and axial drift by manual segmentation of three-dimensional image stacks, which we have used as our control for the true track volume and drift values. Secondly, track parameters were calculated via the single outline, maximum-intensity projections method described above. (This method can only document drift angles in two dimensions). Finally, the frustum method was used to calculate track parameters in three dimensions. Table 3 lists the results of these measurements and their respective errors. While this is far from a complete sample of tracks, the three-dimensional frustum method appears to be more accurate than the single slice, maximum-intensity projection method in measuring track volume. Additionally, the frustum method is extremely accurate in measuring axial drift. The frustum method is dependent on the existence of three-dimensional data, but it is very fast, and its accuracy will be directly correlated to the number of frustums a given track is divided into. Hundreds of data points could be collected in under an hour, as opposed to full track outlines, which can take days to complete. Furthermore, automated segmentation techniques on frustums could prove easier than traditional slices through tracks, due to the roughly elliptical nature of track cross-sections. We look forward to increasing our suite of frustum measurements in the future.

**Entrance Hole Structure**

　　　As an indicator of initial particle size, no single measurement is as readily relied upon as entrance hole diameter. Burchell et al. (2006) originally derived a relationship between original particle size and entrance hole size from controlled experimental analog shots, and this relationship has been further refined by Hörz et al. (2009). For measurement and analysis, the entrance area should be a focus of primary concern, for it is the initial structure created by the pristine impactor. This work seeks to address two parameters of the entrance hole largely overlooked by other researchers. First is the eccentricity of entrance holes. Especially on smaller tracks, we have observed non-circular entrance holes. Two-dimensional entrance hole size measurements could be off by a large margin, because few impactors are spherical. Iida et al. (2010) reported on Track 100, with a highly elliptical entrance hole (in 3D) and a terminal particle akin to the shape of a sandal. Entrance hole area and shape, measured in 3D, provide a first, best indicator of original impactor size and shape. Track geometries quickly become cylindrical with penetration, and less representative of actual impactor shapes (Iida et al. 2010).

　　　Entrance hole deflection is another track feature that is not commonly discussed in the literature. We define the entrance hole deflection as the distance from the apparent aerogel block surface to the local minimum track diameter chosen as the entrance hole location in the track length reference frame (D). This feature is observable in two-dimensional cross sections. Upon inspection in three dimensions, deflection usually shows a parabolic dish shape, exhibiting remarkable symmetry around the entrance hole. Deflection can be quantified using two parameters: depth and width. At present, it is unclear how this deflection feature is formed, possible explanations include: spallation of the aerogel surface around the impact, aerogel compression around the site of original impact, and aerogel vaporization due to the extreme temperatures produced by initial impact (Coulson 2009). We observe no evidence in SXRF data (due to buildup of trace contaminants in aerogel) for compression of aerogel around track entrance holes. Aerogel tends to fracture along planes in the laboratory, but no such features are associated with deflection. The most likely explanation is vaporization on impact. One way to determine the nature of the material surrounding the entrance hole would be FIB-TEM analysis of aerogel around this structure, to determine structural properties of altered aerogel. First order physics lead us to conclude that the size of the deflection feature should be related to the size and kinetic energy of the original impactor. Measurement of deflection parameters in 3D is unreliable when aerogel surfaces are dirty, or on keystones that have been re-mounted in kapton envelopes. Figure 10 plots the entrance hole deflection vs. track length, for the four tracks where this feature has been reliably measured. Optimally one would plot entrance hole deflection vs. impactor mass, but significant errors could propagate due to mass lost in the vaporization of volatile material. Figure 10 suggests a direct relationship between track length and entrance hole deflection, and this relationship should provide a means for estimating original impactor mass. A larger sample set, along with fiducial measurements from analog shots, is needed for a better calibration of this relationship. We expect to report on such measurements in the next year.

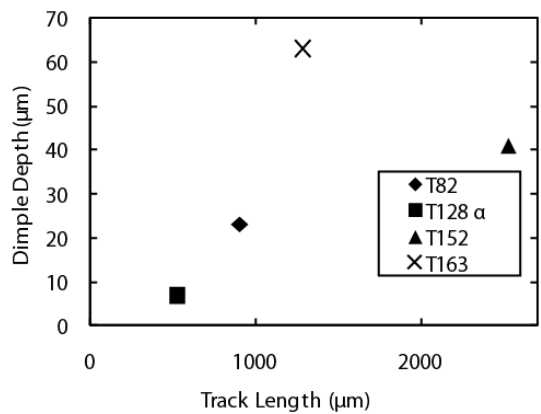

**Figure 10:** Track length vs. Entrance hole depression depth. All accurate measurements are plotted here. Tracks with keystone damage or without accurate measurements are omitted.

**Original Impactor Properties**

Using all of our data in combination we can begin to estimate original impactor size and mass. These calculations are preliminary and illustrative in that we anticipate higher accuracy in imaging with an upgraded instrument, and further refinements in SXRF analysis. We require fully segmented three-dimensional data, in addition to full track, high resolution SXRF maps. Tracks 152 and 163 are the only two that meet these criteria. The first track we discuss is Track 152. Visual inspection of the entrance area of track 152 as well as analysis of the cross sectional area in the stylus (the ~2000 μm above TP1 in Fig. 7) indicates that the impacting particle was not spherical, but had an eccentricity ~1.31. Using the experimentally calibrated equations of Burchell et al. (2008), we can estimate the original impactor size to be an ellipsoid 10.2 μm x 9.25 μm, where the third axis is estimated to be the average of the two, at ~9.78 μm. Integrating the volume of this ellipsoid, we estimate an impactor volume of 3840 μm$^3$. The bulk compositions of elements with Z > 16 derived from SXRF results without normalization to CI are similar to those reported for other tracks (Zolensky et al. 2006). Whole track integrations of SXRF data indicate a total collected mass of ~7.2 ng, for the assumption of CI bulk composition (Flynn et al. 2006; Ishii et al. 2008; Lanzirotti et al. 2008). We can calculate an original impactor density from these measurements, and come to a value of ~1.9 g/cm$^3$. A 7.2 ng impactor with a CI composition and density should occupy a volume of 3400 μm$^3$ (Britt and Consolmagno 2000). Our calculation suggests that ~440 μm$^3$ of the original impactor was porosity beyond that of CI meteorites, or was present as material that was vaporized and lost during track formation.

Similar measurements can be used to make comparable estimates for a bulbous type B track, such as Track 163. In this case it is once again apparent that the impactor shape was not spherical, and the measured eccentricity of the track profile drifts significantly, ranging from 0.4 – 0.94. Using the relationship of Burchell et al. (2008), we can estimate the original particle dimensions to be 13.4 μm x 12.0 μm x their average, resulting in an ellipsoid with a volume of 1079 μm$^3$. Integrating whole track SXRF data results in a CI normalized mass of 1.2 ng. A 1.2 ng CI impactor (density 2.12 g/cm$^3$) should occupy a volume of 567 μm$^3$, about half the estimated impactor volume. This is to be expected in bulbous tracks, where the ratio of volatile to lithic content is expected to be higher (Trigo-Rodríguez et al. 2008). These are first order estimates that illustrate how the properties of impactors may be derived by combined analysis of the datasets presented above. While our measurements of entry hole geometry and track morphology are probably the most accurate available, stricter calibration against controlled experimental measurements (e.g., Burchell et al. 2008) will allow for more accurate estimations of original impactor properties. While our SXRF calculations follow established practices, the calculation of true impactor mass distributed along tracks remains a difficult problem. Improvements in treatment of SXRF data should be possible, to allow improved accuracy of such calculations.

## CONCLUSIONS

LSCM has provided unmatched, highest-resolution imagery of cometary tracks of a variety of sizes. Analysis yields several important track parameters including volume of the cavity along the track, and measurements of deflection of the aerogel surface around the impact hole. SXRF yields an estimate of the total mass of cometary material deposited in each track, as a function of track depth. These datasets can be combined to recover original properties of impacting grains, such as pore space and volatile content, lost in the capture process. So far, we have been able to derive first order estimates consistent with the original properties expected for a carrot-shaped track impactor (Track 152), and a bulbous track impactor (Track 163). As our image quality improves and we begin to differentiate track components, with improved SXRF data analysis, and with tighter constraints from experimental analog shots, we should be able to more accurately determine the physical properties of the original cometary grains that impacted the Stardust aerogel collectors.

*Acknowledgements:*
We thank D. Frank (JSC) and A. Westphal (Berkeley) for their sample preparation efforts, and R. Rudolph and J. Thostenson (AMNH) for LSCM assistance. We also thank B. Garelick and M. Hammer for their efforts in data processing. Thanks also to M. Newville and S.

Sutton for advice and assistance at the APS. Use of the APS is supported by the U.S. Department of Energy, Office of Science, Office of Basic Energy Sciences, Contract No. DE-AC02-06CH11357. Research was funded by NASA LARS through NNG06GE42G (DSE). Editorial handling by John Bradley is appreciated.